\documentclass[12pt]{iopart}
\usepackage{graphicx}
\usepackage{amssymb}
\usepackage{iopams}
\usepackage{dcolumn}
\usepackage{bm}
\usepackage{epsfig}
\usepackage{setspace}
\usepackage[caption=false]{subfig}
\usepackage{xcolor}
\begin{document}

\title{Net-baryon number variance and kurtosis within nonequilibrium chiral fluid dynamics}

\author{Christoph Herold$^1$, Marlene Nahrgang$^2$, Yupeng Yan$^1$, Chinorat Kobdaj$^1$}

\address{$^1$ School of Physics, Suranaree University of Technology, 111 University Avenue, Nakhon Ratchasima 30000, Thailand}
\address{$^2$ Department of Physics, Duke University, Durham, NC 27708, USA}

\ead{herold@g.sut.ac.th}
\vspace{10pt}
\begin{indented}
\item[]July 2014
\end{indented}

\begin{abstract}

We study the variance and kurtosis of the net-baryon number in a fluid dynamical model for heavy-ion collisions.
It is based on an effective chiral model with dilatons for the strong coupling regime of QCD. Taking into account spinodal instabilities, 
we demonstrate that this model exhibits a 
diverging quark number susceptibility and kurtosis all along the spinodal lines of the first-order phase transition, with a 
change of universality class at the critical end point. During the (3+1) dimensional expansion of a hot and dense fireball, 
instabilities are created by fluctuations in the explicitly propagated chiral and dilaton field. We find a clear enhancement of 
event-by-event fluctuations of the baryon number at the critical point and first-order phase transition in comparison with an evolution through the 
crossover region. 

\end{abstract}

\pacs{5.75.-q, 47.75.+f, 11.30.Qc, 24.60.Ky, 25.75.Nq}

\vspace{2pc}
\noindent{\it Keywords}: relativistic heavy-ion collisions, relativistic fluid dynamics, dynamic symmetry breaking, fluctuations at first-order phase transitions

\submitto{\JPG}

\maketitle
 

\section{Introduction}

Disclosing the phase structure of strongly interacting matter is one of the major motivations in the study of finite-temperature quantum chromodynamics (QCD).
The volume independence of susceptibilities from lattice QCD data confirms a crossover chiral and deconfinement transition at small values of the 
baryochemical potential $\mu_{\rm B}$ \cite{Aoki:2006we,Borsanyi:2010bp}. In the regime of large densities, studies of effective models like the 
linear sigma or Nambu-Jona-Lasinio (NJL) model suggest a first-order phase transition and a critical end point (CEP) 
\cite{Scavenius:2000qd,Schaefer:2007pw,Fukushima:2008wg,Herbst:2010rf}. This can be further supported by investigations within the approach of 
Dyson-Schwinger equations \cite{Fischer:2014ata}. Due to various approximations and limitations of all these methods there is, however, no agreement 
about the location of the CEP and 
transition line. During the RHIC beam energy scan, STAR has recently reported measurement 
of directed flow \cite{Adamczyk:2014ipa}, which might be interpreted as experimental evidence for a first-order phase transition.

In order to determine the transition temperature and chemical potential experimentally, quantities which may signal a chiral 
phase transition are required. Of special interest in this context are susceptibilities of conserved charges like the net-baryon number or 
electric charge, who have been shown to display a peak at a crossover and first-order phase transition and a divergence at a CEP 
\cite{Redlich:2006rf,Schaefer:2006ds}. Such fluctuations have been proposed as experimental observable for the detection of a CEP in a heavy-ion 
collision \cite{Stephanov:1998dy,Stephanov:1999zu}. However, measurements in the NA49 experiment could hardly find any non-monotonic behavior 
\cite{Alt:2008ab,Anticic:2008aa}. It has later been shown that higher moments or cumulants and their ratios are even more 
sensitive to a critical structure \cite{Stephanov:2008qz}. Of particular interest here is the kurtosis, a volume-independent quantity which becomes 
negative on the crossover side of the CEP \cite{Skokov:2010uh,Stephanov:2011pb}. The beam-energy scan carried out by the STAR collaboration 
was able to find deviations of the kurtosis from hadron resonance gas and UrQMD transport model calculations \cite{Aggarwal:2010wy,Adamczyk:2013dal}. 
As an alternative to the measurement of fluctuations, it has been shown in \cite{Asakawa:2008ti} that the ratio of antiprotons to 
protons is sensitive to the presence of a CEP due to a focusing of the isentropic trajectories.

It is important to note that all predictions have been made under the assumption that the phase transition takes place in equilibrium, resulting 
both in divergent fluctuations at a CEP and finite ones at the first-order transition. 
However, the system produced in a heavy-ion collision is rapidly expanding and cooling which makes it inevitable to consider dynamical effects. 
Besides the finite system size, critical slowing down is expected to influence the dynamics near a CEP. This has been demonstrated phenomenologically 
in \cite{Berdnikov:1999ph} and within a nonequilibrium fluid dynamical model in \cite{Nahrgang:2011mv,Herold:2013bi}. At a dynamical first-order phase transition, 
spinodal instabilities play a crucial role. Including them in the NJL model, the authors in \cite{Sasaki:2007db} succeeded to demonstrate 
how the quark number susceptibility diverges all along the isothermal spinodals. These divergences result from the convex structure of the pressure 
and the presence of a mechanically instable region. Consequently, one would expect large fluctuations not only at a CEP but also, and possibly 
stronger, at a first-order phase transition. Here, the fast collective expansion of matter produced after the collision of two nuclei should 
lead to the formation of a supercooled phase \cite{Csernai:1995zn,Zabrodin:1998dk,Keranen:2002sw,Nahrgang:2011vn}. If nucleation times are large, this phase will spinodally decompose 
\cite{Mishustin:1998eq,Randrup:2009gp,Randrup:2010ax}, leading to domain formation in the order parameter fields \cite{Herold:2013bi} 
and non-uniform structures like droplets in the baryon density, driven by pressure gradients. The subsequent hadronization 
of such droplets would result in non-statistical multiplicity fluctuations and an enhancement of higher flow harmonics 
\cite{Steinheimer:2012gc,Herold201414}. In order to draw final conclusions from the experimental data, models for a dynamical phase transition 
including critical behavior and also finite size and time effects are required. This would also allow predictions for future 
experiments at FAIR \cite{Friman:2011zz} and NICA \cite{nica:whitepaper} which will cover the region of high densities in the QCD phase diagram. 

In this article we present a study of event-by-event fluctuations from a fully dynamical model of heavy-ion collisions. Starting from 
a linear sigma model with dilatons \cite{Sasaki:2011sd}, we couple a fluid of quarks and gluons to the explicit propagation of the sigma field
as the chiral order parameter and the dilaton representing a gluon condensate. Such an ansatz has been pursued for the first time in 
\cite{Mishustin:1998yc}, where the production and collapse of vacuum bubbles was observed during the expansion of the chiral fluid. We go beyond this 
study by augmenting the classical Euler-Lagrange equations for the sigma field with terms for dissipation and noise, considering the proper and 
full nonequilibrium dynamics from the interaction of the locally thermalized fluid with the out-of-equilibrium evolution of the field. The 
corresponding Langevin equation has been derived selfconsistently in \cite{Nahrgang:2011mg}. In former dynamical studies the gluons were included 
on the basis of the Polyakov loop \cite{Herold:2013bi}, a static quantity defined in Euclidean space-time. In contrast to this, the dilaton field 
has two advantages: First, it comes with a kinetic term in the Lagrangian, making its dynamics straightforward to derive. Second, the problem 
of negative pressures at a first-order phase transition in the Polyakov loop model \cite{Steinheimer:2012gc} can be avoided. 

We begin with a description of the model and the equations of motion in Sec.~\ref{sec:model}, followed by the calculation of the quark number 
susceptibility and kurtosis at a nonequilibrium first-order phase transition in the regime of low temperatures in Sec.~\ref{sec:suscep}. 
In Sec.~\ref{sec:trajectories}, we focus on the impact of a nonequilibrium evolution on fluctuation observables by determining the variance and 
kurtosis of the net-baryon number distribution in an event-by-event study. 
We conclude with a summary and outlook in Sec.~\ref{sec:summary}.

\section{Nonequilibrium chiral fluid dynamics}
\label{sec:model}

We provide a dynamical nonequilibrium model based on a linear sigma model with a dilaton field \cite{Sasaki:2011sd}, for which the Lagrangian density reads
\begin{eqnarray}
\label{eq:Lagrangian}
{\cal L}&=&\overline{q}\left(i \gamma^\mu \partial_\mu-g_{\rm q} \sigma\right)q + \frac{1}{2}\left(\partial_\mu\sigma\right)^2 
+ \frac{1}{2}\left(\partial_\mu\chi\right)^2 + {\cal L}_A- U_{\sigma}-U_{\chi}~, \\
\label{eq:LagrangianA}
 {\cal L}_A&=&-\frac{1}{4}A_{\mu\nu}A^{\mu\nu}+\frac{1}{2}g_A^2\left(\frac{\chi}{\chi_0}\right)^2 A_\mu A^\mu~, \\
U_{\sigma}&=&\frac{\lambda^2}{4}\left[\sigma^2-f_{\pi}^2\left(\frac{\chi}{\chi_0}\right)^2\right]^2-h\left(\frac{\chi}{\chi_0}\right)^2\sigma~, \\
U_{\chi}&=&\frac{1}{4}B\left(\frac{\chi}{\chi_0}\right)^4\left[\ln\left(\frac{\chi}{\chi_0}\right)^4-1\right]~.
\end{eqnarray}
In addition to the usual linear sigma model which describes the melting of the chiral condensate $\sigma\sim\langle\bar q q\rangle$ at high 
temperatures or net-baryon densities, it includes a dilaton or glueball field which we may identify with the gluon condensate 
$\langle A_{\mu\nu}A^{\mu\nu}\rangle$. The term ${\cal L}_A$ in the Lagrangian stands
for a constituent gluon field $A_\mu$ which acquires mass from the nonvanishing expectation value of the gluon condensate $\langle\chi\rangle$. Its 
field strength tensor is defined as $A_{\mu\nu}=\partial_\mu A_\nu-\partial_\nu A_\mu$. We consider only the light quarks $q=(u,d)$ in our present 
study.

The model captures essential features of QCD in the strong coupling regime, the spontaneous breakdown of 
chiral symmetry and the trace anomaly. An alternative approach including the Polyakov loop as the thermal Wilson line over the color-electric 
field $A_0$ has been pursued in \cite{Herold:2013bi}. However, this Polyakov-quark-meson (PQM) model \cite{Schaefer:2004en} yields negative 
values of the pressure at a first-order phase transition with spinodal instabilities \cite{Steinheimer:2013xxa}.

The masses of both the constituent quarks and gluons are dynamically generated via $m_{\rm q}=g_{\rm q}\sigma$ and $m_A=g_A\chi/\chi_0$. From this the 
coupling constants are fixed by reproducing the vacuum nucleon and glueball masses $3m_{\rm q}=m_N=940$~MeV and $2m_A=m_G=1.7$~GeV. The term $h=f_\pi m_\pi^2$ with 
the pion decay constant $f_\pi=93$~MeV and the pion mass $m_\pi=138$~MeV explicitly breaks chiral symmetry. In vacuum, the energy density equals 
$B/4=0.76~\mbox{GeV}/\mbox{fm}^3$ which determines the bag constant B. The dimensionful parameter $\chi_0$ is obtained from setting the vacuum glueball 
mass $m_G$ equal to the second derivative of the potential $U_{\chi}$. The self-coupling of the chiral field can be evaluated as 
$\lambda^2=\frac{m_\pi^2-m_\sigma^2}{2f_\pi^2}$ and thus depends on the vacuum sigma mass. As shown in \cite{Sasaki:2011sd}, a value of 
$m_\sigma=900$~MeV yields a reasonable behavior of the gluon condensate around the chiral transition in comparison with lattice QCD data. 
At nonzero baryochemical potential, lattice calculations have to rely on sophisticated methods to circumvent the infamous sign problem via 
reweighting \cite{Fodor:2001pe} or an imaginary chemical potential \cite{deForcrand:2002ci}. Obtained results are however still inconclusive about both 
the existence and position of a CEP. The corresponding phase diagram of our model has a chiral critical point at temperature $T_{\rm CEP}=89$~MeV 
and quark chemical potential $\mu_{\rm CEP}=329$~MeV 
with an adjacent first-order phase transition line. The critical chemical potential here is rather large compared with recent results from Dyson-Schwinger 
equations \cite{Fischer:2014ata} or a chiral quark-hadron model \cite{Dexheimer:2009hi} which predict values of $\mu_{\rm CEP}$ below $200$~MeV. 
On the other hand, a similarly large $\mu_{\rm CEP}$ and correspondingly small $T_{\rm CEP}$ are predicted by a PQM model supplemented with 
fluctuations \cite{Herbst:2010rf} or the mean-field NJL model \cite{Scavenius:2000qd}. The linear sigma model with dilatons and also the PQM model 
approach the Stefan-Boltzmann limit at large temperatures which the standard linear sigma model fails to reproduce \cite{Sasaki:2011sd}. At 
large chemical potentials, however, the equation of state for the PQM model predicts negative values of the pressure in the region with spinodal 
instabilities. In contrast to that, the equation of state is well-behaved for the model with dilatons.
Above the temperature of $250$~MeV, scale symmetry is 
restored in the dilaton model, here it predicts a strong first-order phase transition which is not seen in lattice QCD. This defect is 
nevertheless neglectable for our studies as we are not going to probe the regime of deconfined gluons but focus on the chiral transition only. 

Within the mean-field approximation, the effective thermodynamic potential is obtained by a path integration over the quark and gluon fields 
\begin{equation}
 V_{\rm eff}=\Omega_{q\bar q}+\Omega_{A}+U_{\sigma}+U_{\chi}+\Omega_0~.
\end{equation}
The quark and gluon contributions can be evaluated as
\begin{eqnarray} 
\Omega_{\rm q\bar q}&=&-2 N_f N_c T\int\frac{\mathrm d^3 p}{(2\pi)^3} \left\{\ln\left[1+\mathrm e^{-\frac{E_{\rm q}-\mu}{T}}\right]+\ln\left[1+\mathrm e^{-\frac{E_{\rm q}+\mu}{T}}\right]\right\}~, \\
\Omega_{A}&=&2 (N_c^2-1) T\int\frac{\mathrm d^3 p}{(2\pi)^3} \left\{\ln\left[1-\mathrm e^{-\frac{E_A}{T}}\right]\right\}~,
\end{eqnarray}
with the quasiparticle energies $E_{\rm q}=\sqrt{p^2+m_{\rm q}^2}$ and $E_A=\sqrt{p^2+m_A^2}$, respectively. Here and in the following, 
$\mu=\mu_{\rm B}$ denotes the quark chemical potential. A constant term $\Omega_0$ is added to ensure 
zero potential and pressure in vacuum. 

Having integrated out the quark and gluon degrees of freedom, we treat them as an ideal fluid to mimic the quark-gluon plasma at high energy 
densities. The fluid is described by the energy-momentum tensor $T^{\mu\nu}=(e+p)u^\mu u^\nu-p g^{\mu\nu}$ and the quark number current 
$N_{\rm q}^\mu=n_{\rm q} u^\mu$. The pressure is given by
\begin{equation}
\label{eq:pressure}
 p = -\Omega_{q\bar q}-\Omega_{A}~,
\end{equation}
from where energy and quark density are obtained via the standard thermodynamic relations $e=T\partial p/\partial T +\mu n_{\rm q}-p$ and 
$n_{\rm q}=\partial p/\partial \mu$.
A full nonequilibrium dynamics for the coupled system of the sigma field and quarks has been derived in \cite{Nahrgang:2011mg} from the 
two-particle irreducible effective action. We adopt the result for the extended model with gluons and dilatons, the Langevin equation 
of motion for the sigma field reading
\begin{equation}
\label{eq:eomsigma}
 \partial_\mu\partial^\mu\sigma+\eta_{\sigma}\partial_t \sigma+\frac{\delta V_{\rm eff}}{\delta\sigma}=\xi_{\sigma}~.
\end{equation}
In addition to the classical Euler-Lagrange equation it contains a damping coefficient $\eta_\sigma$ and a stochastic noise field $\xi_{\sigma}$ 
describing dissipation and noise in the thermalized heat bath of quarks and gluons. Physically, the damping occurs due to the decay of one sigma 
into a quark-antiquark pair. As the sigma meson becomes light around the phase transition, also $\eta_\sigma$ decreases and eventually vanishes 
at the critical point. Its explicit form is given by 
\begin{equation}
\label{eq:dampingcoeff}
  \eta_{\sigma}=\frac{12 g^2}{\pi}\left[1-2n_{\rm F}\left(\frac{m_\sigma}{2}\right)\right]\frac{1}{m_\sigma^2}\left(\frac{m_\sigma^2}{4}-m_{\rm q}^2\right)^{3/2}~.
\end{equation}
We work in the approximation of Gaussian white noise, with the noise field correlator
\begin{equation}
\label{eq:dissfluctsigma}
 \langle\xi_{\sigma}(t,\vec x)\xi_{\sigma}(t',\vec x')\rangle_\xi=\delta(\vec x-\vec x')\delta(t-t')m_\sigma\eta_{\sigma}\coth\left(\frac{m_\sigma}{2T}\right)~.
\end{equation}
Similar to the sigma in the quark fluid, one might expect the dilaton in the gluonic medium to be damped, too. The corresponding process would be 
the emission of two gluons $\chi\rightarrow\chi+\rm g+g$, according to Eq. (\ref{eq:LagrangianA}). However, as the dilaton mass is of the order 
of two times the in-medium mass of the gluons, this process is kinematically forbidden. We therefore apply the classical equation of motion 
to the dilaton field
\begin{equation}
\label{eq:eomchi}
 \partial_\mu\partial^\mu\chi+\frac{\delta V_{\rm eff}}{\delta\chi}=0~.
\end{equation}
Conservation of the overall energy-momentum as well as the baryon number are ensured by the fluid dynamical equations
\begin{eqnarray}
\label{eq:fluidT}
\partial_\mu T^{\mu\nu}&=&-\partial_\mu\left(T_\sigma^{\mu\nu}+T_\chi^{\mu\nu}\right)~,\\
\label{eq:fluidN}
\partial_\mu N_{\rm q}^{\mu}&=&0~.
\end{eqnarray}
Mainly due to the aforementioned dissipation, the fields lose energy to the fluid which is transferred via the source terms 
in Eq. (\ref{eq:fluidT}). This effect will be especially significant at a first-order phase transition, during the formation of a 
supercooled phase and its subsequent decay \cite{Herold:2013bi,Nahrgang:2011vn}. Since the evolution of the sigma field is stochastic in nature, 
also the fluid dynamical equations (\ref{eq:fluidT}), (\ref{eq:fluidN}) are becoming stochastic via the coupling to the source term. Fluctuating fluid dynamics has 
recently attracted attention in the context of heavy-ion collisions \cite{Kapusta:2011gt}.
Finally, our set of equations is closed by a nonequilibrium equation of state, where the pressure explicitly depends on the local values of 
the fields $\sigma$ and $\chi$, cf. Eq. (\ref{eq:pressure}).

\section{Susceptibilities in the spinodal region}
\label{sec:suscep}

\begin{figure}[t]
\centering
    \subfloat[\label{fig:suscep}]{
    \centering
    \includegraphics[scale=0.61,angle=270]{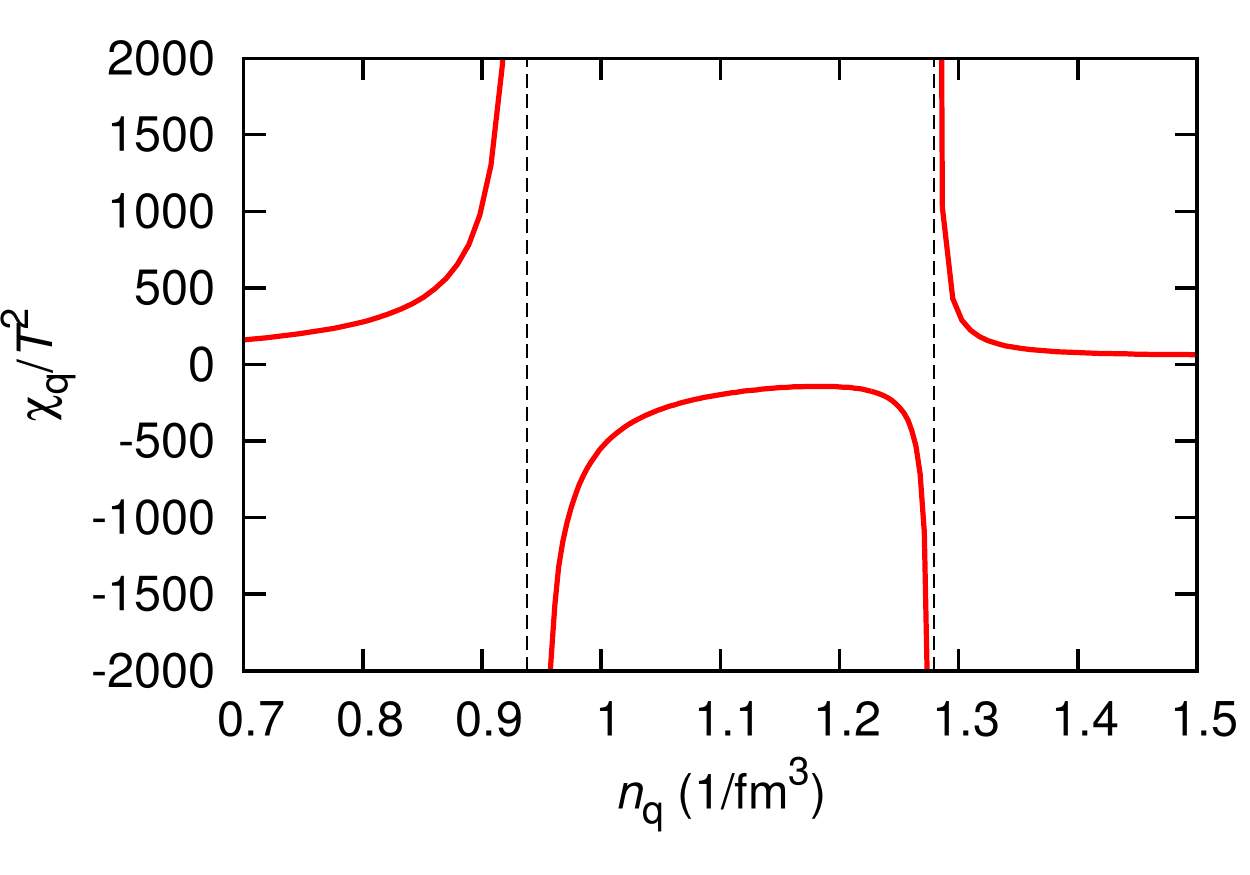}
    }
  \hfill
    \subfloat[\label{fig:kurtosis}]{
    \centering
    \includegraphics[scale=0.61,angle=270]{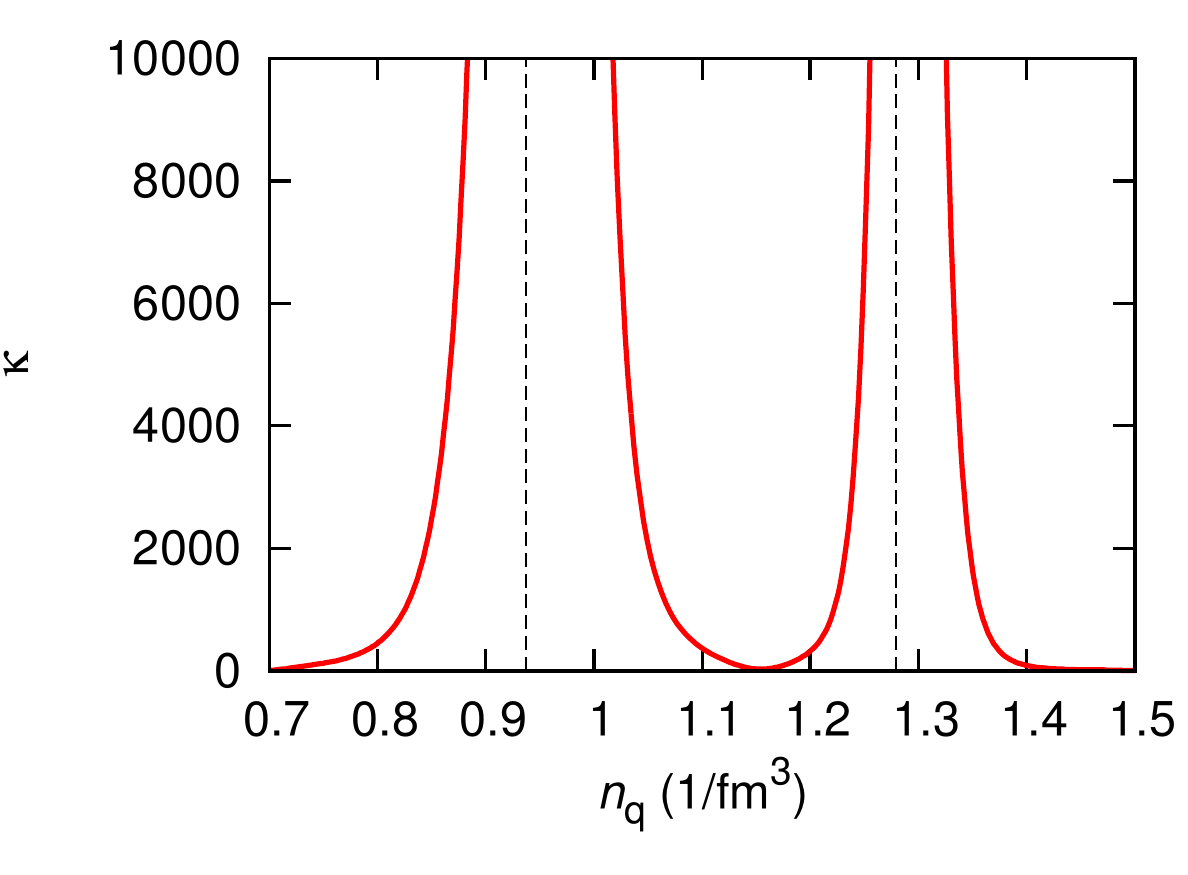}
    }
\caption[Quark number susceptibility and kurtosis]{Quark number susceptibility \subref{fig:suscep} and kurtosis \subref{fig:kurtosis} 
for a nonequilibrium first-order phase transition at $T=40$~MeV.}
\label{fig:susceptibilities}
\end{figure}

Of particular interest for the detection and localization of the chiral phase transition are fluctuations of conserved charges like the net-baryon 
number. From effective models, one can calculate susceptibilities which describe fluctuations in the quark number density as response to changes 
in the chemical potential. They are in general defined as
\begin{equation}
 c_n = \frac{\partial^n(p/T^4)}{\partial(\mu/T)^n}~.
\end{equation}
Here we focus on two coefficients, namely $c_2$, which is proportional to the quark number susceptibility $\chi_{\rm q}$ and the variance of fluctuations 
$\sigma^2$, and the kurtosis $\kappa$, given by the 
ratio of the fourth to the second coefficient. They are related to fluctuations in the quark number $N_{\rm q}$ as
\begin{eqnarray}
 c_2&=&\frac{\chi_{\rm q}}{T^2}=\frac{\sigma^2}{V T^3}=\frac{1}{V T^3}\langle\delta N_{\rm q}^2\rangle~, \\
 \frac{c_4}{c_2}&=&\kappa=\frac{\langle\delta N_{\rm q}^4\rangle}{\langle\delta N_{\rm q}^2\rangle}-3\langle\delta N_{\rm q}^2\rangle~,
\end{eqnarray}
with $\delta N_{\rm q}=N_{\rm q}-\langle N_{\rm q}\rangle$, the deviation from the ensemble average of the quark number distribution. Note that for $\kappa$ the volume and 
temperature dependence cancel to leading order. In \cite{Stephanov:1999zu} it has been 
shown that event-by-event fluctuations like $\langle\delta N_{\rm q}^2\rangle$ diverge at a critical point. The non-monotonic 
behavior of fluctuations in heavy-ion collisions as a function of beam energy was proposed as an experimental signal. It was later shown in \cite{Sasaki:2007db} within 
the NJL model that susceptibilities also diverge along the spinodal lines of the first-order phase transition. This requires that nonequilibrium
effects, i. e. spinodal instabilities, are taken into account. The presence of a mechanically instable region with $\partial p/\partial n_{\rm q} <0$ then 
consequently leads to diverging susceptibilities. For the sigma model with dilatons, we calculate both the quark number susceptibility 
and the kurtosis as a function of density for fixed temperature $T=40$~MeV, where the model exhibits a first-order phase transition, see Fig. 
\ref{fig:susceptibilities}. In the left plot, the susceptibility is shown as a function of the quark density. Similar to the result from the NJL 
model, we see divergences at the isothermal spinodal points. Inside the spinodal region, the susceptibility becomes negative due to instabilities. 
On the right hand side we show the kurtosis in the same density range. Here we also find strong divergences when crossing the spinodal lines. 
Interestingly, $\kappa$ remains positive even inside the coexistence region. On the other hand, it is known that the kurtosis becomes negative 
when approaching the critical end point from the crossover side \cite{Skokov:2010uh}. 

\begin{figure}[t]
\centering
    \subfloat[\label{fig:expsusc}]{
    \centering
    \includegraphics[scale=0.62,angle=270]{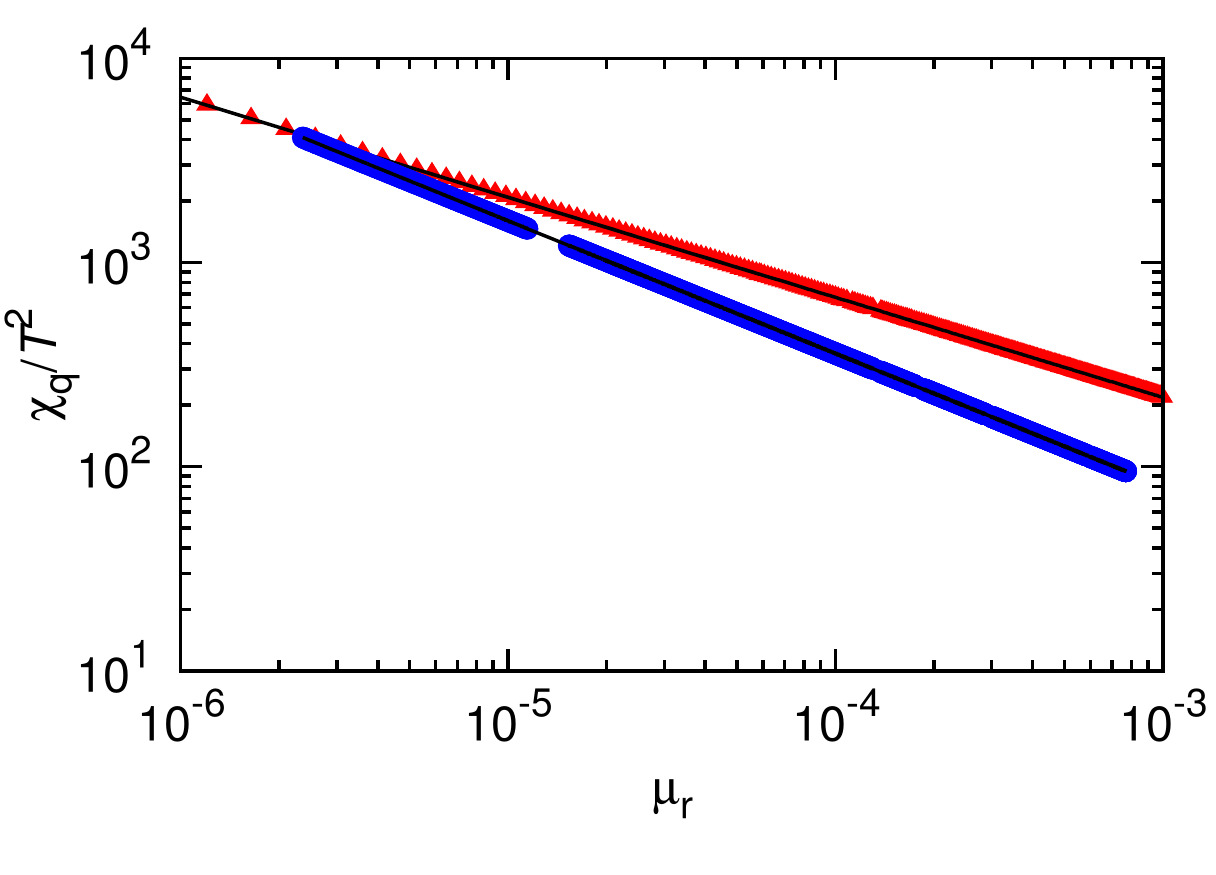}
    }
  \hfill
    \subfloat[\label{fig:expkurtosis}]{
    \centering
    \includegraphics[scale=0.62,angle=270]{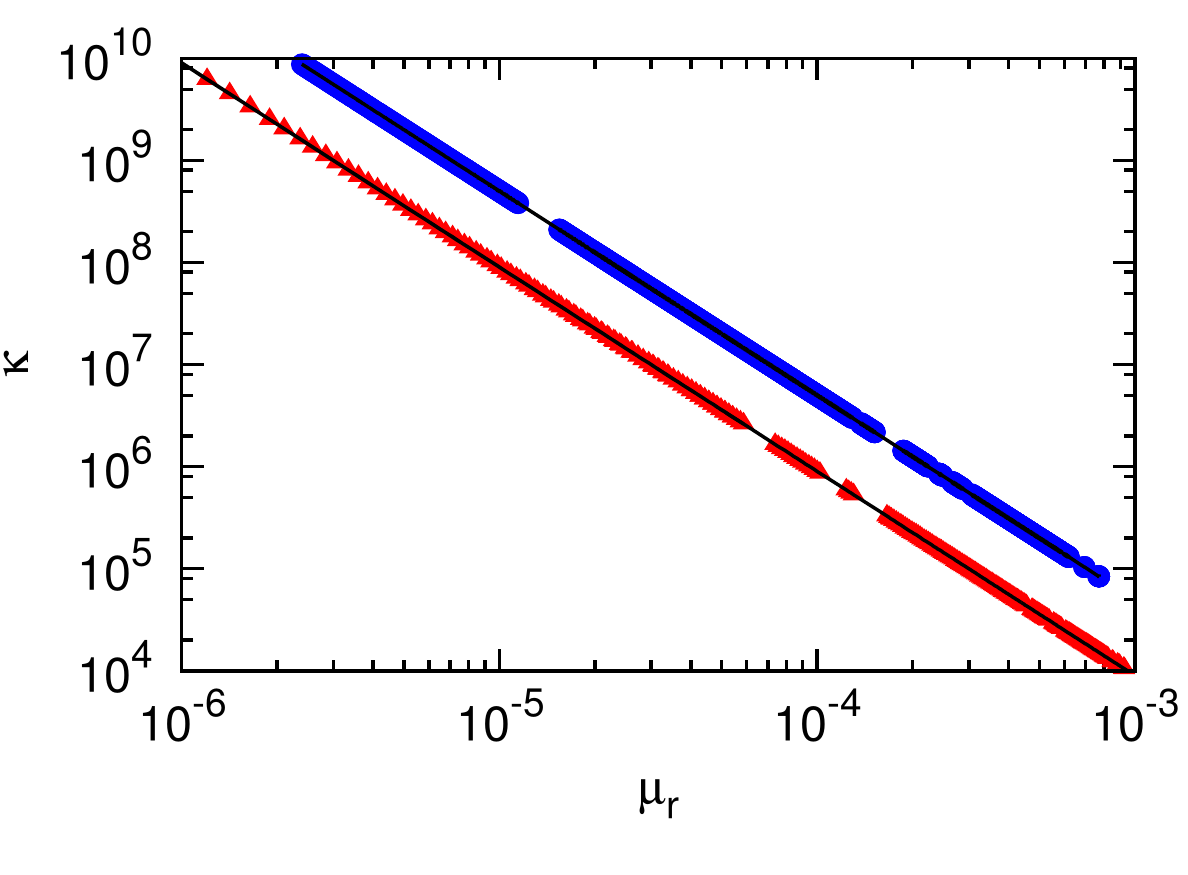}
    }
\caption[Quark number susceptibility and kurtosis]{Quark number susceptibility \subref{fig:expsusc} and kurtosis \subref{fig:expkurtosis} 
as function of the reduced chemical potential near the critical point (circles) and first-order phase transition (triangles) for fixed 
temperature $T=40$~MeV.}
\label{fig:critical}
\end{figure}

Further insight into the nature of these fluctuations can be gained by determining the critical exponents which control the strength of the 
divergences at the CEP or spinodals, respectively. In the vicinity of the singularity, the behavior of susceptibility and kurtosis may be 
described by a power law of the form 
\begin{eqnarray}
 \chi_{\rm q} &\sim & (\mu-\mu_0)^{-\gamma}~, \\
 \kappa &\sim & (\mu-\mu_0)^{-\zeta}~.
\label{eq:critexp}
\end{eqnarray}
We calculate the critical exponents $\gamma$ and $\zeta$ both analytically and numerically. The chiral transition may be described by a Ginzburg-Landau 
effective theory around the CEP and the spinodal lines \cite{Sasaki:2007qh}. At both points, the first and second derivatives of the effective potential 
vanish. Around a zero of the second derivative, it can be expanded in terms of $\delta\sigma=\sigma-\sigma_0$
\begin{equation}
 V_{\rm eff}=a_0 +a_1\delta\sigma+a_2\delta\sigma^2+a_3\delta\sigma^3+a_4\delta\sigma^4~.
\end{equation}
At $\delta\sigma=0$ we have $a_1=a_2=0$, so for these coefficients the relation $a_1=b_1(\mu-\mu_0)$ and $a_2=b_2(\mu-\mu_0)$ holds for small values of 
$\delta\sigma$. From $\partial V_{\rm eff} /\partial \sigma=0$ we obtain $a_1+a_2\delta\sigma+a_3\delta\sigma^2+a_4\delta\sigma^3 =0$, and we 
can assume that $\delta\sigma\sim (\mu-\mu_0)^\alpha$ with $0<\alpha<1$. At the spinodal point, the leading term reads $a_1+a_3\delta\sigma^2=0$, 
giving $\alpha=1/2$ and consequently $\gamma=1/2$. For a CEP, we also have $a_3=0$, so from $a_1+a_4\delta\sigma^3=0$ we end up with 
$\alpha=1/3$ and $\gamma=2/3$. For the fourth generalized susceptibility $c_4$, we can immediately state the coefficients to be $5/2$ for 
the CEP and $8/3$ for the spinodal. As the kurtosis is proportional to the ratio of the fourth to the second derivative of the effective potential 
with respect to $\mu$, we get $\zeta=2$ for both cases. The same values for $\gamma$ and $\zeta$ have been found within a numerical 
analysis fitting the forms in Eqs. (\ref{eq:critexp}) to the numerically determined susceptibility and kurtosis. Fig. \ref{fig:critical} shows 
the analytical and the numerical results versus the reduced chemical potential $\mu_{\rm r}=(\mu-\mu_0)/\mu_0$ for illustration. 

The critical properties for a CEP and an isothermal spinodal point are different due to a change in the universality class, indicating 
different critical behavior and strength of divergences. The same result and 
exponents have been found for a chiral NJL model with finite current quark masses \cite{Sasaki:2007qh}. The critical exponents of the kurtosis are naturally in agreement for 
both types of transition and are found to be equal to $2$. 

\section{Nonequilibrium enhancement of fluctuation signals}
\label{sec:trajectories}

\begin{figure}[t]
\centering
    \includegraphics[scale=0.7]{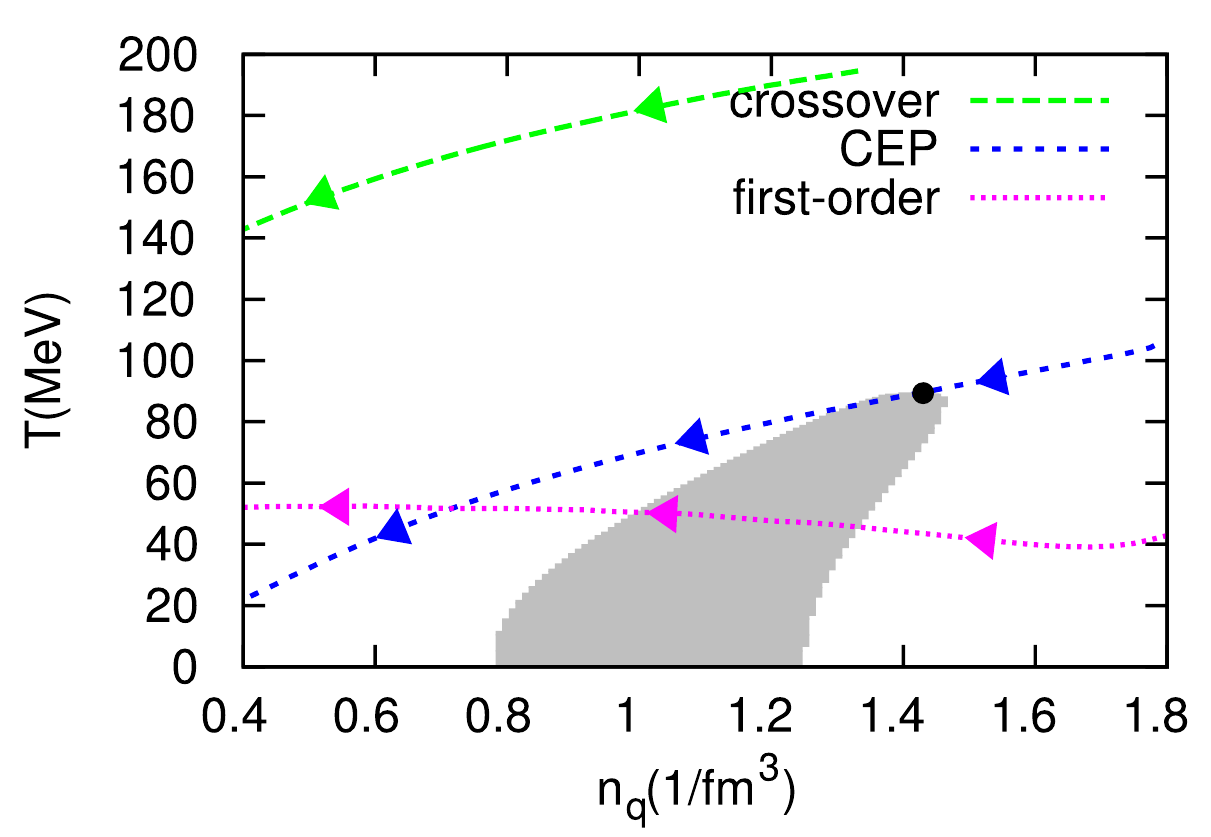}
\caption[Trajectories]{Trajectories for a crossover, critical and spinodal transition. The gray area depicts the spinodal region and the 
black circle indicates the position of the CEP. Arrows show the direction of the evolution. }
\label{fig:traj}
\end{figure}

The results from the previous section indicate that in dynamical systems as they are created in heavy-ion collisions, fluctuation signals may not 
only be enhanced in the vicinity of a critical point, but also in the spinodal region. This might provide us with a more applicable means to 
investigate the QCD phase structure as the region around the CEP with enhanced susceptibility is small 
\cite{Schaefer:2006ds,Kunihiro:1991qu,Hatta:2002sj} and subject to finite size and time effects that limit the growth of fluctuations. 
We may therefore expect larger fluctuations at a first-order phase transition than at a CEP. We test this assumption within the 
nonequilibrium chiral fluid dynamics model introduced in Sec.~\ref{sec:model}. We initialize a spherical droplet of quark-gluon plasma 
by defining an initial temperature and quark chemical potential, with a Woods-Saxon distribution to ensure a smooth transition to the vacuum 
at the edges. 
Then the fields are initialized with their respective equilibrium distribution, assuming Gaussian fluctuations around the thermal 
expectation values $\langle\sigma\rangle$ and $\langle\chi\rangle$, and finally we calculate the fluid dynamical quantities out of the values 
for $T$, $\mu$, $\sigma$ and $\chi$. Fluctuations in the initial conditions have only minor influence on the evolution as these are quickly 
washed out by the damping and superposed by the stochastic noise. By choosing appropriate initial values for $T$ and $\mu$, 
we are able to observe the expansion and cooling through the crossover, critical and spinodal region. The total quark number is in 
each case fixed to $N_{\rm q}=67$. In Fig.~\ref{fig:traj}, we show event-averaged trajectories for the three scenarios in the $T$-$n_{\rm q}$-plane. 
The values of the density and temperature in a single event are obtained from an averaging over a central volume. 
Each single cell follows of course its individual path through the phase diagram, and plotting all of them would yield a blob moving from 
higher to lower densities. The volume-averaged trajectories differ from event to event due to different noise configurations. 
The curves start on the right side 
proceeding to lower density on the left. Interestingly, we see that the first-order curve shows a slightly increasing temperature at intermediate 
densities between $1.6/\mbox{fm}^3$ to $0.6/\mbox{fm}^3$. This is a result of the reheating effect that occurs after the decomposition of a 
supercooled phase and is typical for a first-order phase transition. It has already been found in earlier works of nonequilibrium fluid dynamical 
models \cite{Herold:2013bi,Nahrgang:2011vn}. Note that this is a purely dynamical effect which causes the trajectory to even cross the CEP curve, 
where the temperature decreases monotonically. It also implies that for this curve there is some significant deviation from the equilibrium 
trajectory along the corresponding isentrope. 
This crossing was also observed for the earlier used PQM model \cite{Herold201414} if the curves were plotted in the 
$T$-$n_{\rm q}$-plane and the initial conditions were chosen close enough to each other. In the present model the initial conditions for a CEP and 
first-order transiton are inevitably closer as $T_{\rm CEP}$ is comparatively low. 
For the first-order phase transition we also find bubbles created through spinodal decomposition, 
as has been reported in earlier fluid dynamical studies \cite{Herold:2013bi,Herold201414}. However, in the present model 
these high-density droplets are not stable but start to decay after traversing the spinodal region. This is a direct consequence of the now strictly
positive pressure as has been pointed out already in the introduction.

\begin{figure}[t]
\centering
    \subfloat[\label{fig:ebesuscep}]{
    \centering
    \includegraphics[scale=0.7,angle=270]{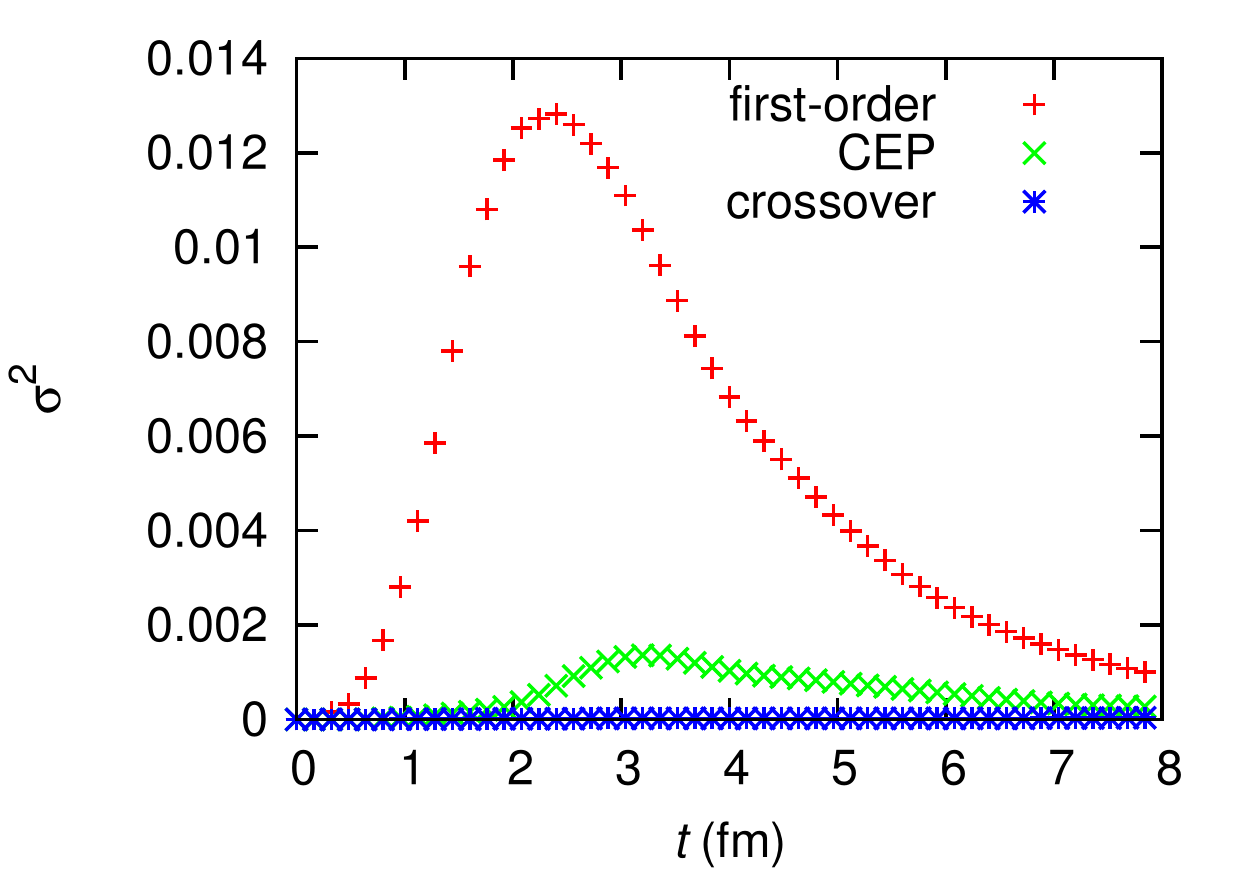}
    }
  \hfill
    \subfloat[\label{fig:ebekurtosis}]{
    \centering
    \includegraphics[scale=0.7,angle=270]{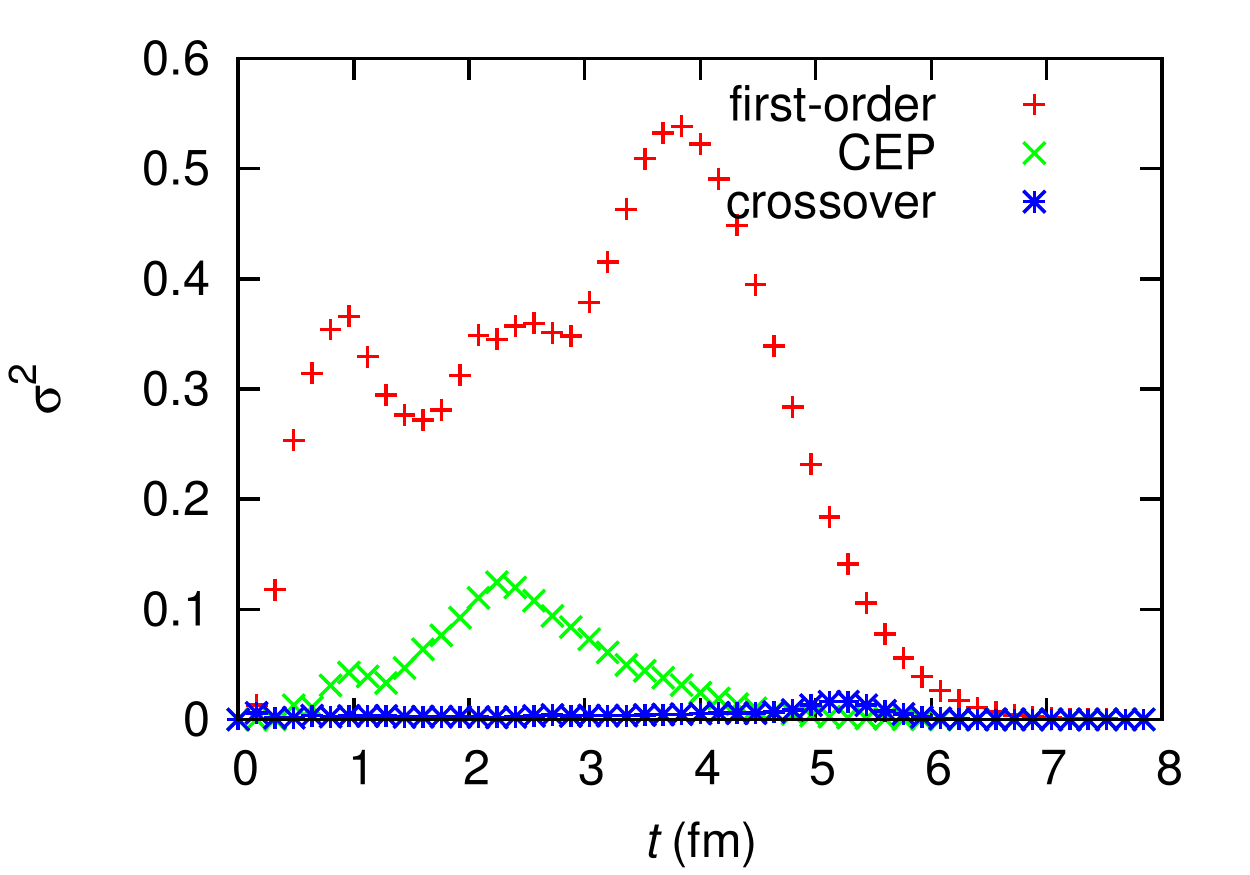}
    }
\caption[Event-by-event fluctuations]{Variance of the net-baryon number from the fluid dynamical evolution in a fixed volume \subref{fig:ebesuscep} and with rapidity and transverse momentum cut \subref{fig:ebekurtosis}.}
\label{fig:ebe}
\end{figure}

We show the evolution of event-by-event fluctuations in the net-baryon number $N_{\rm B}=3 N_{\rm q}$ corresponding to these trajectories as a function of time in 
Fig.~\ref{fig:ebe}. The baryon number is extracted directly from the fluid dynamical density. As a conserved quantity, it will fluctuate 
only slightly when including a freeze-out and hadronic interactions in the final state \cite{Koch:1987}. We use two different methods to determine 
$N_{\rm B}$: First, within a fixed volume in the center of the collision, with 
an extension of $10$~fm in $x$-direction and $1$~fm each in $y$- and $z$-direction. Second, by limiting the region of acceptance via 
rapidity to $|y|<0.5$ and transverse momentum density to $100 \mbox{ MeV/fm}^3<p_T<500 \mbox{ MeV/fm}^3$ as it was performed in recent measurements at STAR
\cite{Adamczyk:2014ipa}. Both methods yield qualitatively similar results. The variance of fluctuations is enhanced at a CEP in comparison 
with a crossover transition and even more, by a factor of $5$ to $6$, at a first-order phase transition.  As the variance depends on the volume, 
we observe clear differences in the scales of the two plots. Furthermore, the volume strongly varies when applying a constant 
rapidity and momentum cut, therefore a more irregular structure in the time dependence can be found for that case. 

\begin{figure}[t]
\centering
    \subfloat[\label{fig:ebesuscep2}]{
    \centering
    \includegraphics[scale=0.7,angle=270]{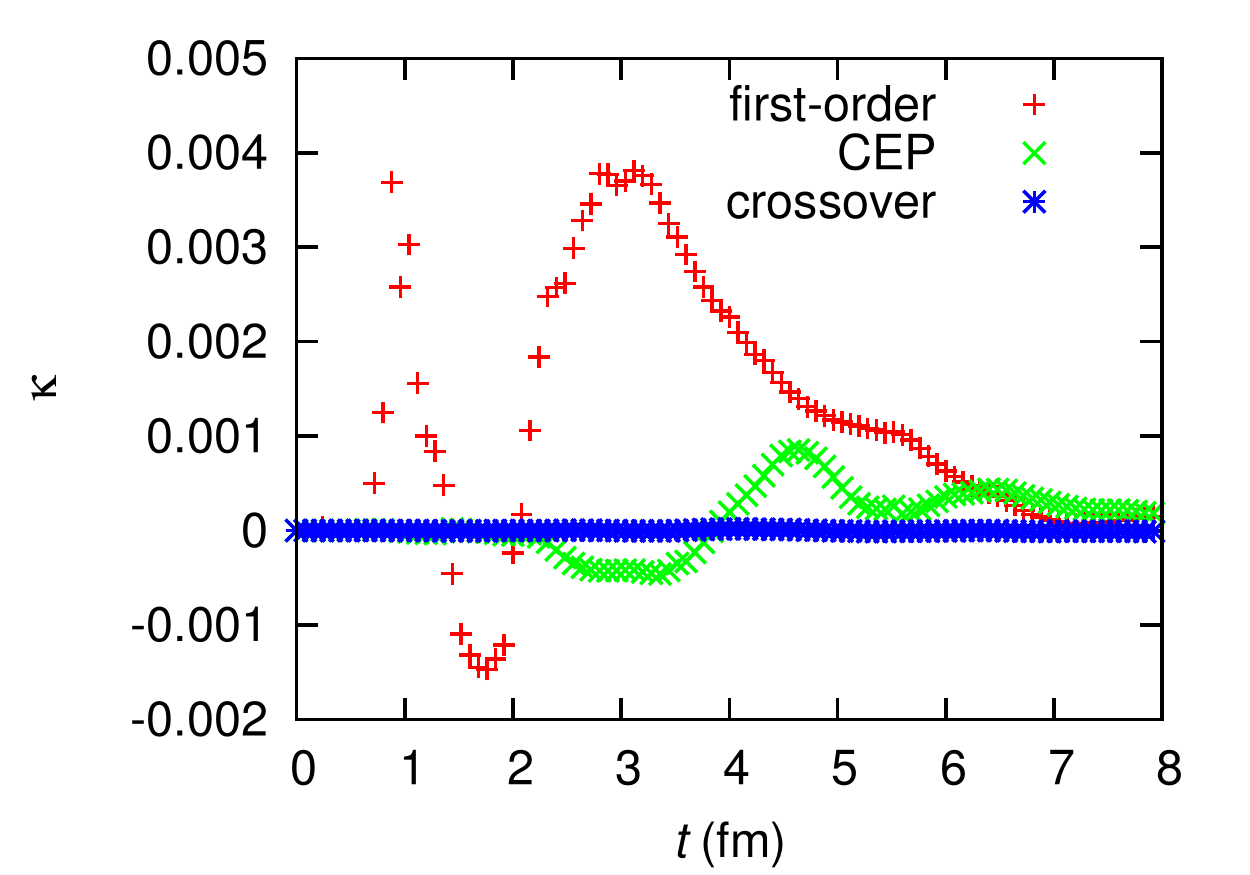}
    }
  \hfill
    \subfloat[\label{fig:ebekurtosis2}]{
    \centering
    \includegraphics[scale=0.7,angle=270]{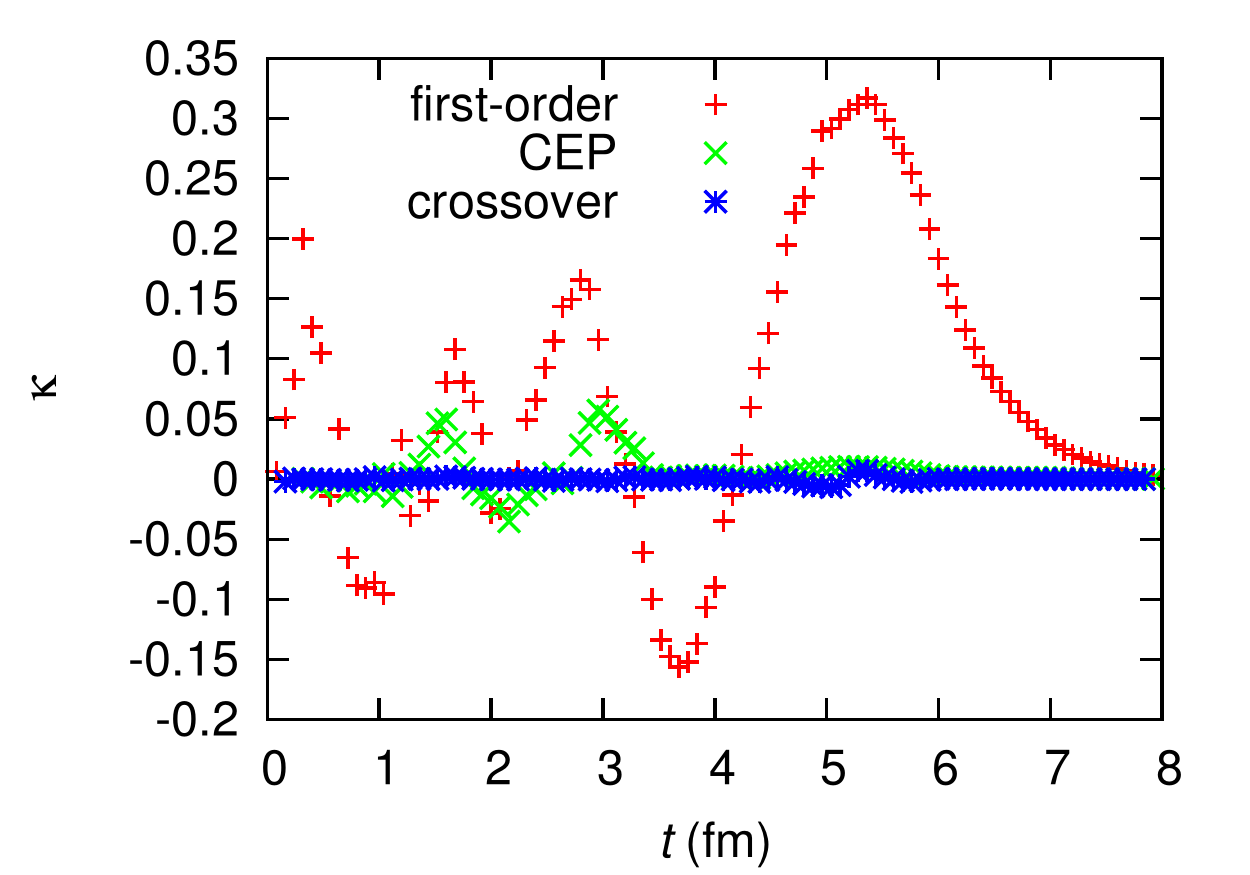}
    }
\caption[Event-by-event fluctuations]{Kurtosis of the net-baryon number from the fluid dynamical evolution in a fixed volume \subref{fig:ebesuscep2} and with rapidity and transverse momentum cut \subref{fig:ebekurtosis2}.}
\label{fig:ebe2}
\end{figure}

For the kurtosis in Fig.~\ref{fig:ebe2}, we also find that the crossover transition produces values close to zero, while a clear enhancement can 
be found at the CEP. Again, the largest fluctuations occur for a first-order phase transition. Remarkably, at CEP and first-order transition, 
both positive and negative values of $\kappa$ occur during the evolution, confirming the assumption of critical behavior in the spinodal region. 
Although the kurtosis is not dependent on the volume, we find its values to be an order of magnitude higher when using a rapidity and momentum cut 
than in the case of a fixed test volume. This can be explained considering the overall conservation of quark or baryon number in the system under 
consideration. In contrast to that, the baryon number is only on average conserved in a grand canonical ensemble which is used for effective model 
or lattice QCD calculations. As shown in \cite{Bleicher:2000ek,urqmdkurtosis,Bzdak:2012an}, this global conservation significantly effects ratios of cumulants, making them 
dependent on the fraction of measured to total baryons. 

At this point we should note that it is nontrivial to draw the connection between the time evolution of the variance and kurtosis and 
event-by-event fluctuations from experiment, which are supposed to be emitted over a hypersurface of constant energy density or temperature. 
The latter can only be measured after freeze-out and therefore a signal from the phase transition 
can only be extracted if the chemical freeze-out temperature is close to the temperature of hadronization. Otherwise, the fluctuations may have 
been washed out after passing the phase transition. Finally, one needs to relate the baryon number fluctuations to quantities that are actually 
measured, like fluctuations in the proton number \cite{Kitazawa:2012at}, and consider the evolution of these fluctuations in the hadronic phase 
\cite{Kitazawa:2013bta}.

\section{Summary and Outlook}
\label{sec:summary}

We have investigated baryon number fluctuations within a chiral model with dilatons from two different approaches: First, through the 
calculation of susceptibilities, where we went beyond standard thermodynamics by including spinodal instabilities. We were able 
to show that both the quark number susceptibility and kurtosis diverge at the CEP and spinodal lines. The singularity in the susceptibility at the 
spinodals becomes suddenly stronger at the CEP, indicated by a larger critical exponent. The implications of such a behavior for experiment 
are strong enhancements of event-by-event fluctuations in the net-baryon number, which we investigated in a second dynamical approach. 
Propagating the chiral field and the dilaton explicitly on a locally thermalized background of quarks and gluons, we simulated the expansion 
of the hot and dense plasma created in a heavy-ion collision. We extracted the variance and kurtosis of the net-baryon number. Both are stronger 
at a CEP in comparison with a crossover scenario, and even more enhanced when the system evolves through the spinodal region of the first-order 
phase transition. 

In the future we are going to include hadronic degrees of freedom for a more realistic description of the chirally broken and confined phase. 
It is furthermore necessary to study particle distributions from a freeze-out or hadronic afterburner. This would also allow us to study 
the momentum anisotropy and the effect of the phase transition on flow. For the determination of susceptibilities, it would be interesting 
to include quantum or thermal fluctuations and study their effect on the critical properties near the CEP and first-order phase transition.

\section*{Acknowledgements}

This work is funded by Suranaree University of Technology (SUT) and CHE-NRU (NV.12/2557) project. The authors thank Igor Mishustin and Chihiro Sasaki 
for fruitful discussions and Dirk Rischke for providing the SHASTA code that was used for the fluid dynamical simulation. M. N. acknowledges support 
from the U.S. Department of Energy under grant DE-FG02-05ER41367 and a fellowship within the Postdoc-Program of the German 
Academic Exchange Service (DAAD).
The computing resources have been provided by the National e-Science Infrastructure 
Consortium of Thailand, the Center for Computer Services at SUT and the Frankfurt Center for Scientific Computing.

\section*{References}
\bibliographystyle{unsrt}
\bibliography{mybib}

\end{document}